\documentclass[preprint,showpacs,preprintnumbers,amsmath,amssymb]{revtex4}

\usepackage{graphicx}
\usepackage{dcolumn}
\usepackage{bm}

\begin{document}


\title{Vanishing conductivity of quantum solitons in polyacetylene}

\author{Leonardo Mondaini}
 \email{leo.mondaini@pq.cnpq.br}
\affiliation{Centro Federal de Educa\c c\~ao Tecnol\'ogica Celso Suckow da Fonseca, UnED Nova Friburgo, Nova Friburgo RJ 28600-000, Brazil}

\author{E. C. Marino}
 \email{marino@if.ufrj.br}
\affiliation{Instituto de F\'{\i}sica, Universidade Federal do Rio de Janeiro,
Cx. Postal 68528, Rio de Janeiro RJ 21941-972, Brazil}

\author{A. A. Schmidt}
 \email{alex@lana.ccne.ufsm.br}
\affiliation{Departamento de Matem\'atica, Universidade Federal de Santa Maria, Santa Maria RS 97105-900, Brazil}


\begin{abstract}
Quantum solitons or polarons are supposed to play a crucial role in
the electric conductivity of polyacetylene, in the intermediate
doping regime. We present an exact fully quantized calculation of
the quantum soliton conductivity in polyacetylene and show that it
vanishes exactly. This is obtained by applying a general method of
soliton quantization, based on order-disorder duality, to a $Z(2)$-symmetric complex extension of the TLM dimerization effective field theory. We show that, in this theory, polyacetylene solitons
are sine-Gordon solitons in the phase of the complex field.
\end{abstract}


\maketitle


\section{Introduction}

The discovery of a tremendous increase in the electrical
conductivity of {\it trans}-polyacetylene, when doped either with
halogens or alkalis, \cite{doping} was a breakthrough of far
reaching consequences in physics and chemistry. The fact that the
{\it trans}-isomer occurs in two degenerate species opens the
possibility of occurrence of soliton defects interconnecting them.
This fact unfolded an enormous range of possibilities
interconnecting many areas, including mathematics, theoretical and
experimental physics. The subject has remained on the focus of
interest until recently \cite{r1}. It actually happens that such
topological excitations are produced in the process of doping
\cite{SSH1,SSH2}. Pure polyacetylene has one active $\pi$-electron
per site and is a Peierls insulator, due to the electron-lattice
interaction. It has been found that in the presence of a soliton,
treated at the classical level, electron states are created in the
middle of the gap, hence it is energetically favorable for the extra
doped electrons to create solitons and occupy the midgap states
rather than going into the conduction band.

From the very beginning the existence of three different doping
regimes became clear. Firstly, for low doping concentrations (small
compared to 1\%) the above picture of classical solitons works very
well. The solitons are pinned by the dopant atoms, which create them
with the corresponding midgap states. The conductivity is thermally
activated, corresponding to a transition from the midgap to the
conduction band states and can be understood quite similarly to the
conductivity in semiconductors. Secondly, for high concentration of
dopants (5\% to 10\%) the conduction regime is clearly metallic,
with an unfilled conduction band, and can be thereby understood.
Thirdly, there is an intermediate regime of dopant concentration, of
the order of 1\%, in which none of the previous models work. In this
regime, the solitons become dynamic carriers of charge and a full
quantum treatment of these excitations becomes unavoidable since the
soliton mass is of the same order of the electron mass
\cite{heeger}.

The purpose of this work is to apply a general method of
quantization of soliton excitations \cite{sq2}, in order to describe
the conductivity of polyacetylene in the intermediate regime. In
order to do that, however, the following obstacle must be removed.
From the mathematical point of view, the soliton is a topologically
nontrivial configuration of the dimerization - or phonon - field,
which describes the lattice degrees of freedom. This, of course, is
a real field, whose effective potential has $Z(2)$-symmetry and two
degenerate minima that correspond to the two species of {\it
trans}-polyacetylene. The above method of soliton quantization,
however, only applies to complex fields, in the case of a
multiplicative symmetry such as $Z(2)$ \cite{sq2}.

Therefore, in order to describe the quantum solitons of the system
and specifically their role in the electric conduction in the
intermediate doping regime, we propose a $Z(2)$-symmetric complex
extension of the effective potential for the dimerization field.
This has the same topological properties of the former and,
consequently should not alter substantially the soliton physics.

In Section 2, we describe the method of soliton quantization in a
theory of a complex scalar field with $Z(N)$ symmetry and show that
the soliton excitations of it are sine-Gordon (SG) solitons in the
phase of the complex scalar field. In Section 3, we propose the
$N=2$ version of this theory as the complex extension of the
effective theory for the dimerization field in polyacetylene. In
Section 4, we derive an exact series expression for the quantum
soliton current-current correlation function and, out of it, obtain
the soliton conductivity. We show that this exactly vanishes. In
conclusion, we have an exact demonstration at a full quantum level,
that dynamic solitons are actually not the carriers of charge in
polyacetylene. Rather, polarons, which are basically
soliton-antisoliton bound states should be responsible for the
transport of charge in the intermediate doping regime of
polyacetylene.

\section{Quantum phase solitons in theories with a Z(N) symmetry}

We start by considering the following theory describing a complex
scalar field in (1+1)-dimensions,
\begin{equation}
\mathcal{L}=\partial_\mu\phi^\ast\partial^\mu\phi+\gamma\left({\phi^\ast}^N
+\phi^N\right)-\eta\left(\phi^\ast\phi\right)^M, \label{5.1.1}
\end{equation}
where $N$ and $M$ are integers and $\gamma$ and $\eta$ are real
parameters. This is invariant under the $Z(N)$ transformation:
$\phi(x,t)\rightarrow e^{i\frac{2\pi}{N}}\phi(x,t)$. The choice
$\gamma > 0$, implies the spontaneous breakdown of the $Z(N)$
symmetry. In this case, the theory will have degenerate vacua and
soliton excitations. A full quantum theory of these solitons was
developed in \cite{sq2,ms,mbs}. This includes an explicit expression
for the soliton creation operator, namely
\begin{equation}
\mu(x)=\exp\left\{-\frac{2\pi}{N}\int_{x,C}^{\infty}d\xi_\nu
\epsilon^{\mu\nu}
\phi^\ast(\xi){\stackrel{\leftrightarrow}{\partial}}_\mu
\phi(\xi)\right\}, \label{1.2.2.8}
\end{equation}
and a general expression for its local Euclidean correlation
functions \cite{sq2,ms,mbs},
\begin{equation}
\langle\mu(x)\mu^\dagger(y)\rangle = \mathcal{N}\int
\mathcal{D}\phi^\ast \mathcal{D}\phi \exp\left\{-\int
d^2z\left[\left(D_\mu\phi\right)^\ast
\left(D_\mu\phi\right)+V\left(\phi^\ast,\phi\right)\right]\right\},\label{1.2.2.3}
\end{equation}
where
\begin{equation}
D_\mu=\partial_\mu-i\alpha A_\mu, \ \ \ \ \ \ \ \ \ \ \ \
A_\mu(z,C)=\int_{x,C}^{y}d\xi_\nu
\epsilon^{\mu\nu}\delta^2(z-\xi).\label{1.2.2.4}
\end{equation}
In the above expression, $V$ is the potential of an arbitrary
Lagrangian and the integral is taken along an arbitrary curve $C$,
connecting $x$ and $y$. It can be shown, however, that
(\ref{1.2.2.3}) is independent of the chosen curve.

We are going to show in what follows that these quantum solitons may
be identified with SG quantum solitons in the phase of the field
$\phi$.

Using the polar representation for $\phi$, namely,
$\phi(x,t)=\rho(x,t) e^{i\theta(x,t)}$, where $\rho$ and $\theta$
are real fields, we can rewrite the above Lagrangian as
\begin{equation}
\mathcal{L}=\partial_\mu\rho\partial^\mu\rho+\rho^2\partial_\mu\theta\partial^\mu\theta+2\gamma\rho^N\cos
N\theta-\eta\rho^{2M} . \label{5.1.3}
\end{equation}

In what follows, we will be interested in the topological properties
of the theory. As we shall argue, these are not affected by $\rho$
fluctuations, hence, from now on we will make the constant $\rho$
approximation,
\begin{equation}
\rho(x,t)=\rho_0, \ \ \ \ \ \ \rho_0 \,\ \textrm{constant}.
\label{5.1.4}
\end{equation}
Using this in (\ref{5.1.3}) we get
\begin{equation}
\mathcal{L}=\rho_0^2\partial_\mu\theta\partial^\mu\theta+2\gamma\rho_0^N\cos
N\theta-\eta\rho_0^{2M}, \label{5.1.5}
\end{equation}
which is a SG Lagrangian in $\theta$.

We conclude that, in the constant-$\rho$ approximation, the theories
given by (\ref{5.1.1}) will present SG solitons in the phase of the
complex scalar field $\phi$. The corresponding topological current
will be
\begin{equation}
J^\mu=\epsilon^{\mu\nu}\partial_\nu \theta, \label{5.1.5a}
\end{equation}
which is associated to the topological charge operator
\begin{equation}
\mathcal{Q}=\int_{-\infty}^\infty dx' \, J^0 = \int_{-\infty}^\infty
dx' \,
\partial_{x'} \theta(x',t)= \theta(+\infty,t)- \theta(-\infty,t). \label{1.2.3.2}
\end{equation}

We see that topological properties are related to large $\theta$
fluctuations and, therefore, the constant $\rho$ approximation
should not interfere in such properties.

In order to explicitly confirm the fact that the soliton operators
introduced in \cite{sq2,ms,mbs} are indeed creation operators of
quantum solitons in the phase of the field $\phi$, let us explicitly
evaluate the commutation relation between the quantum soliton
creation operator and the topological charge.


Observe that this is nothing but the Mandelstam creation operator of
quantum solitons in the SG model \cite{mand}, as it should. Then,
using canonical commutation relations we readily find
\begin{eqnarray}
\left[\mathcal{Q},\mu\right] =
\left\{-i\frac{2\pi}{N}\int_{-\infty}^\infty dx' \, \partial_{x'}
\int_{x}^{\infty}d\xi_1 \left[\theta(x',t)\ \ ,\ \
\pi_\theta(\xi_1,t)\right]\right\}\mu
\nonumber \\ = \frac{2\pi}{N}\mu.
\label{1.2.3.4}
\end{eqnarray}

Eq. (\ref{1.2.3.4}) implies that the operator $\mu$ creates
eigenstates of the topological charge $\mathcal{Q}$, with eigenvalue
$2\pi/N$, thus proving that the quantum solitons occurring in the
theory described by (\ref{5.1.1}) are indeed {\it phase solitons}.
Correlation functions of these quantum soliton excitations have been
calculated elsewhere \cite{MM1,MM2,MM3}. In Section 4, we will show that
the relevant quantum correlators for the calculation of the
conductivity will be the soliton current-current correlators.

\section{The case N=2: a model for polyacetylene}

In this section, we are going to propose a phenomenological theory
for polyacetylene that will enable us to compute quantum soliton
correlation functions and in particular the correlation functions of
quantum solitonic current operators. As we shall see the standard
field theory model for this polymer unfortunately does not allow the
application of the method of soliton quantization described in the
previous section. For this reason, we will propose a
phenomenological alternative.

Polyacetylene is described by the Su-Schrieffer-Heeger (SSH) model
\cite{SSH1,SSH2}, whose field theory version is the
Takayama-Lin-Liu-Maki (TLM) model \cite{TLM}, described by the
Hamiltonian
\begin{eqnarray}
H_{TLM}=\int dx\,\Psi_s^\dag(x)\left[-i\hbar
v_F\sigma_3\partial_x+\Delta(x)\sigma_1\right]\Psi_s(x) \nonumber \\  +(2\pi\hbar
v_F\lambda)^{-1}\int
dx\,\left[\dot{\Delta}^2(x)/\Omega_0^2+\Delta^2(x)\right].
\label{4.2.6}
\end{eqnarray}

In the above expression, $\Psi_s(x)$ is a two-component Dirac
fermion field, associated to the $\pi$-electrons and $\Delta(x)$ is
a real scalar field - the dimerization field - associated to the
lattice degrees of freedom, namely, the phonons. Furthermore,
$\sigma_i$ are the Pauli matrices, $v_F$ is the Fermi velocity,
$\lambda$ is the dimensionless electron-phonon coupling constant,
and $\Omega_0$ is the bare optical-phonon  frequency.

Integrating over the fermion field in the previous expression, we
obtain an effective theory for the phonon field $\Delta(x)$, whose
potential is a Z(2) symmetric double-well
\cite{campbell1,campbell2,grossneveu}. The two degenerate minima of
this potential correspond to the two degenerate dimerizations of
{\it trans}-polyacetylene. The equivalent calculation has also been
performed within the SSH model \cite{SSH1,SSH2}, also leading to a
degenerate double-well effective potential for the dimerization
variable.

The double-well potential for the effective dimerization field
implies the existence of soliton excitations in the theory. It has
been shown that these, indeed, are introduced by doping
polyacetylene with halogen or alkali atoms \cite{doping}.

As we have argued in the Introduction, in the regime of intermediate
doping, a full quantum treatment of the soliton excitations is
unavoidable. Hence, one would be naturally inclined to use the
method of soliton quantization for theories with a $Z(N)$ symmetry,
described in the previous section, for the case $N=2$. Nevertheless,
one immediately realizes that the method is not applicable for a
real field such as the dimerization field $\Delta(x)$. Indeed, for a
real field the exponent of the soliton operator (\ref{1.2.2.8})
vanishes, making $\mu$ trivial and the soliton correlator
(\ref{1.2.2.3}) no longer makes sense, since we cannot couple the
external field $A_\mu$ to a real field. Furthermore, for a real
field, $\theta =0$, hence the topological current and the respective
topological charge cannot be defined as in (\ref{5.1.5a}) and
(\ref{1.2.3.2}). The above method of soliton quantization, in the
case of a multiplicative symmetry \cite{sq2}, only applies to
complex fields.

In order to conciliate this fact with the knowledge that the
effective theory for the real $\Delta(x)$-field is a degenerate
double-well with a $Z(2)$ symmetry, we propose a complex extension
$\phi(x)$ of the $\Delta(x)$-field, governed by the Lagrangian
(\ref{5.1.1}) with $N=M=2$. The corresponding potential is
\begin{equation}
V(\phi^\ast,\phi)=-\gamma\left({\phi^\ast}^2
+\phi^2\right)+\eta\left(\phi^\ast\phi\right)^2, \label{5.3.1}
\end{equation}
or, in terms of polar fields,
\begin{equation}
V(\rho,\theta)=-2\gamma\rho^2\cos 2\theta+\eta\rho^4. \label{5.3.2}
\end{equation}

As we can see, there are two degenerate minima at
\begin{equation}
(\rho_0,\theta_0)=
\begin{cases}
(\sqrt{\frac{\gamma}{\eta}},0) \\
(\sqrt{\frac{\gamma}{\eta}},\pi)
\end{cases}
\label{5.3.3}
\end{equation}
where the potential has the value
$V(\rho_0,\theta_0)=-\gamma^2/\eta$. Applying, then, the constant
$\rho$ approximation
\begin{equation}
\rho(x,t)\simeq \rho_0=\sqrt{\frac{\gamma}{\eta}} \label{5.3.4}
\end{equation}
and adding $\gamma^2/\eta$ we get the following SG potential for the
phase field $\theta$,
\begin{equation}
V(\theta)=\frac{2\gamma^2}{\eta}\left(1-\cos 2\theta\right).
\label{5.3.5}
\end{equation}

The associated classical solitonic excitations will be
\begin{equation}
\theta(x)=\pm 2\arctan\exp\left[\sqrt{\gamma}\, (x-x_0)\right],
\label{5.3.7}
\end{equation}
where the plus and minus signs correspond respectively to a soliton
($\theta_s(x-x_0)$) and an anti-soliton in the phase of the
$\phi$-field. We see that
\begin{equation}
\lim_{x\rightarrow -\infty} \theta_s(x-x_0)=0\ \ \ \ \ \
\textrm{and}\ \ \ \ \ \ \lim_{x\rightarrow \infty}
\theta_s(x-x_0)=\pi, \label{5.3.8}
\end{equation}
namely, the phase soliton connects the minima of the potential
(\ref{5.3.2}), when $\rho=\sqrt{\gamma/\eta}$.

The actual potential for the dimerization field $\Delta(x)$ and the
$Z(2)$-symmetric complex potential for the field $\phi$, given by
(\ref{5.3.2}), both possess the same topology, related to the
$Z(2)$-symmetry. It is therefore reasonable to expect the same
topological properties in both theories, especially those concerning
solitons. We may adjust the parameters in such a way that the minima
of the real field potential coincide with those of the complex one.

What we are doing is quite similar to what is done when we use
complex functions in order to describe the EM field. The physical E
and B fields will correspond to the real part thereof. The
dimerization field $\Delta$ of the TLM model is the real part of our
$\phi$. As we know, for the polyacetilene soliton, we have the
$\Delta$-field varying from $- \Delta_0$ to $\Delta_0$ in between
the two minima.

 A SG soliton in theta would have the phase of the $\phi$ field varying from
$\pi$ to $0$, implying $\phi$ would vary between $-\rho_0$ and
$\rho_0$. Thus, identifying $\rho_0$ with $\Delta_0$, we can figure
out the relation between the sine-Gordon solitons of our model and
the polyacetylene solitons: the real part of the complex field
$\phi$ for the configuration having a SG soliton in its phase will
be in the same topological class as the $\phi^4$-like soliton of
polyacetylene. Since for each SG soliton there is a soliton in
polyacetylene we may identify the SG-soliton current with the
polymer soliton current. We are going to use, therefore, this
complex extension in order to study the quantum properties of the
soliton excitations by means of the method of soliton quantization
described in the previous section. We will use, in particular, the
SG soliton current for calculating the conductivity.

\section{The quantum soliton current correlator and conductivity}

In this section, we are going to obtain an exact series expression
for the soliton dc-conductiviy in our model for polyacetylene. For
this purpose, the starting point is the well-known Kubo formula
\cite{mahan},
\begin{equation}
\sigma_s^{ij}=\lim_{\omega\rightarrow 0}\lim_{k\rightarrow
0}\frac{1}{\omega}\textrm{Im}\left[\langle J^i
J^j\rangle_{ret}(\omega,k)\right], \label{5.2.0}
\end{equation}
where $\langle J^i J^j\rangle_{ret}(\omega,k)$ is the retarded,
Minkowski space, correlation function of the spatial component of
the soliton current operator $J^\mu$ given by (\ref{5.1.5a}).

We want to evaluate the above current-current correlator within our
field theory model for polyacetylene. The strategy will be to derive
a generating functional for such current correlators in our field
theory model. For this purpose, we introduce the identity
\begin{eqnarray}
1=\int
\mathcal{D}J^\mu\,\delta\left[J^\mu-\epsilon^{\mu\nu}\partial_\nu\theta\right]
\nonumber \\ =\int\mathcal{D}J^\mu\mathcal{D}\lambda_\mu\,\exp\left\{i\int d^2 z\,
\left(J^\mu-\epsilon^{\mu\nu}\partial_\nu\theta\right)\lambda_\mu\right\},
\label{5.2.2}
\end{eqnarray}
in the Euclidian vacuum functional associated to the Lagrangian
(\ref{5.1.5}), obtaining
\begin{eqnarray}
\mathcal{Z}= \mathcal{Z}_0^{-1}\int
\mathcal{D}J^\mu\mathcal{D}\lambda_\mu\mathcal{D}\theta\,\exp\left\{-\frac{1}{\hbar
v_S}\int d^2 z\,
\left[\rho_0^2\partial_\mu\theta\partial_\mu\theta \nonumber \right .\right . \\ \left .\left .-2\gamma\rho_0^N\cos
N\theta  -i\hbar
v_S\left(J^\mu-\epsilon^{\mu\nu}\partial_\nu\theta\right)\lambda_\mu\right]\right\},
\label{5.2.3}
\end{eqnarray}
where we used expression (\ref{5.1.5a}) for the soliton current.
Notice that in the expression above, we no longer make $\hbar=c=1$.
Actually, since the $\phi$-field theory replaces the effective
theory for the phonon field $\Delta$ in the TLM model, we substitute
$c$ for $v_S$, the speed of sound in the polymer.

We now integrate over $\theta$ and $\lambda_\mu$, thereby obtaining
the partition function expressed as the functional integral of the
exponential of an effective $J^\mu$ action \cite{LM}. The $\theta$
integral may be done by the usual expansion in powers of the cosine
term \cite{limasantos}, or equivalently, in powers of $\rho_0^N$.
The resulting functional integrals, both in $\theta$ and
$\lambda_\mu$ are quadratic and the final result is
\begin{eqnarray}
\mathcal{Z}=
\mathcal{N}\sum_{m=0}^{\infty}\frac{\left(\frac{\gamma\rho_0^N}{\hbar
v_S}\right)^{2m}}{(m!)^2}\int\prod_{i=1}^{2m}d^2z_i\int
\mathcal{D}J^\mu\,\exp\left\{-\frac{1}{2}\int d^2z d^2z'\nonumber \right . \\ \left . \times \, J^\mu(\vec
z)\left[\frac{2\rho_0^2}{\hbar v_S}\delta^{\mu\nu}\,\delta^2(\vec z - \vec
{z'})\right]J^\nu(\vec {z'}) \nonumber \right .\\ \left .+\int
d^2z\,\left[iN\sum_{i=1}^{2m}\lambda_i\epsilon^{\mu\alpha}\partial_\alpha
G(\vec z_i-\vec z)\right]J^\mu(\vec z)\right\} ,
\label{5.2.4}
\end{eqnarray}
where $\lambda_i = 1$ for $1 \leq i \leq m$ and $\lambda_i = - 1$
for $n+1 \leq i \leq 2m$ and $G(\vec z)$ is the Euclidian Green
function of the two-dimensional (2D) free massless scalar theory,
which appears naturally since the expansion in powers of the cosine
term is an expansion around such theory.

We note at this point that, should we integrate the above expression
over $J^\mu$, we would obtain the usual Coulomb gas representation
for the vacuum functional of the SG theory \cite{sgcg}. Conversely,
the expression for the generating functional of current correlators
can be obtained by the usual procedure of adding a linear coupling
with a source $K_\mu$ in the exponent of the integrand in the
previous espression, namely,
\begin{eqnarray}
\mathcal{Z}\left[K_\mu\right]=
\mathcal{N}\sum_{m=0}^{\infty}\frac{\left(\frac{\gamma\rho_0^N}{\hbar
v_S}\right)^{2m}}{(m!)^2}\int\prod_{i=1}^{2m}d^2z_i\int
\mathcal{D}J^\mu\,\exp\left\{-\frac{1}{2}\int d^2z d^2z' \nonumber \right . \\ \left . \times J^\mu(\vec
z)\left[\frac{2\rho_0^2}{\hbar v_S}\delta^{\mu\nu}\, \delta^2(\vec z - \vec
{z'})\right]J^\nu(\vec {z'})\nonumber \right . \\ \left .+\int
d^2z\,\left[iN\sum_{i=1}^{2m}\lambda_i\epsilon^{\mu\alpha}\partial_\alpha
G(\vec z_i-\vec z)+\frac{K_\mu}{\hbar v_S}\right]J^\mu(\vec
z)\right\}.
\label{5.2.5}
\end{eqnarray}

$\mathcal{Z}\left[K_\mu\right]$ is the desired generating functional
of $J^\mu$ correlators. Indeed, we have
\begin{equation}
\langle J^\mu(\vec x)J^\nu(\vec y)\rangle=\frac{(\hbar
v_S)^2}{\mathcal{Z}}\frac{\delta^2
\mathcal{Z}\left[K_\mu\right]}{\delta K_\mu(\vec x)\,\delta
K_\nu(\vec y)}\Biggr\rvert_{K_\mu=0} . \label{5.2.7}
\end{equation}

Now, integrating (\ref{5.2.5}) in $J^\mu$, we get
\begin{eqnarray}
\mathcal{Z}\left[K_\mu\right]=
\sum_{m=0}^{\infty}\frac{\left(\frac{\gamma\rho_0^N}{\hbar
v_S}\right)^{2m}}{(m!)^2}\int\prod_{i=1}^{2m}d^2z_i \nonumber \\ \times \exp\left\{\frac{1}{2}\int
d^2z
d^2z'\,\left[iN\sum_{i=1}^{2m}\lambda_i\epsilon^{\mu\alpha}\partial_\alpha
G(\vec z_i-\vec z)+\frac{K_\mu}{\hbar v_S}\right]\nonumber \right .
\\  \left . \times \left[\frac{\hbar v_S}{2\rho_0^2}
\delta^{\mu\nu}\delta^2(\vec z - \vec
{z'})\right]\left[iN\sum_{j=1}^{2m}\lambda_j\epsilon^{\nu\beta}{\partial_\beta}'
G(\vec z_j-\vec {z'})+\frac{{K_\nu}'}{\hbar v_S}\right]\right\}\nonumber
\\ =\sum_{m=0}^{\infty}\frac{\left(\frac{\gamma\rho_0^N}{\hbar
v_S}\right)^{2m}}{(m!)^2}\int\prod_{i=1}^{2m}d^2z_i \nonumber \\
\times \exp\left\{-\frac{N^2\hbar
v_S}{4\rho_0^2}\sum_{i=1}^{2m}\lambda_i\sum_{j=1}^{2m}\lambda_j
G(\vec z_i-\vec z_j)\right\}\nonumber \\
\times\exp\left\{\frac{1}{4\rho_0^2\hbar v_S}\int d^2z
d^2z'\,K_\mu(\vec z)\delta^2(\vec z-\vec {z'})K_\mu(\vec {z'})\right
. \nonumber \\ \left
.+\frac{iN}{2\rho_0^2}\sum_{i=1}^{2m}\lambda_i\int
d^2z\,\epsilon^{\mu\alpha}\partial_\alpha G(\vec z_i-\vec
z)K_\mu(\vec z)\right\}. \label{5.2.6}
\end{eqnarray}

Evaluating the functional derivatives in (\ref{5.2.7}), we obtain
\begin{eqnarray}
\langle J^\mu(\vec x)J^\nu(\vec
y)\rangle=\mathcal{Z}^{-1}\sum_{m=0}^{\infty}\frac{\left(\frac{\gamma\rho_0^N}{\hbar
v_S}\right)^{2m}}{(m!)^2}\int\prod_{i=1}^{2m}d^2z_i \nonumber \\ \times \exp\left\{-\frac{N^2\hbar
v_S}{4\rho_0^2}\sum_{i=1}^{2m}\lambda_i\sum_{j=1}^{2m}\lambda_j
G(\vec z_i-\vec z_j)\right\}\nonumber\\  \times\left\{\frac{\hbar
v_S}{2\rho_0^2}\delta^{\mu\nu}\delta^2(\vec x-\vec
y)-\frac{N^2(\hbar v_S)^2}{4\rho_0^4}\sum_{j=1}^{2m}\lambda_j G(\vec
z_j)\nonumber\right .\\ \left .\times
\sum_{i=1}^{2m}\lambda_i\left(\delta^{\mu\nu}\partial_\alpha^{(y)}\partial_\alpha^{(x)}-\partial_\mu^{(y)}\partial_\nu^{(x)}\right)
G(\vec z_i +(\vec x-\vec y))\right\},
\label{5.2.8}
\end{eqnarray}
where we have made the shift of variable $\vec z_{i(j)}\rightarrow
\vec z_{i(j)}-\vec x$.

We now perform the Fourier transform in the variable $\vec
\chi\equiv \vec x-\vec y$, arriving at
\begin{eqnarray}
\langle J^\mu J^\nu\rangle(\vec
k)=\mathcal{Z}^{-1}\sum_{m=0}^{\infty}\frac{\left(\frac{\gamma\rho_0^N}{\hbar
v_S}\right)^{2m}}{(m!)^2}\int\prod_{i=1}^{2m}d^2z_i\nonumber \\
\times\exp\left\{-\frac{N^2\hbar
v_S}{4\rho_0^2}\sum_{i=1}^{2m}\lambda_i\sum_{j=1}^{2m}\lambda_j
G(\vec z_i-\vec z_j)\right\}\nonumber \\ \times\left\{\frac{\hbar
v_S}{2\rho_0^2}\delta^{\mu\nu}-\frac{N^2(\hbar
v_S)^2}{4\rho_0^4}\sum_{j=1}^{2m}\lambda_j G(\vec z_j)\right
.\nonumber \\ \times \left.\left\{\sum_{i=1}^{2m}\lambda_i e^{-i\vec
k\cdot\vec z_i}\left(\frac{\delta^{\mu\nu}{\vec k}^2-k^\mu
k^\nu}{{\vec k}^2}\right)\right\}\right \}, \label{5.2.10}
\end{eqnarray}
where  $\vec k\equiv(k,\omega)$ and we used the fact that
\begin{equation}
\int
d^2\chi\,\left(\delta^{\mu\nu}\partial_\alpha^{(y)}\partial_\alpha^{(x)}-\partial_\mu^{(y)}\partial_\nu^{(x)}\right)
G(\vec \chi +\vec z_i)e^{i\vec k\cdot\vec \chi}=e^{-i\vec k\cdot\vec
z_i}\left(\frac{\delta^{\mu\nu}{\vec k}^2-k^\mu k^\nu}{{\vec
k}^2}\right). \label{5.2.11}
\end{equation}

From (\ref{5.2.10}) we can get $\langle J^i
J^j\rangle_{ret}(\omega,k)$, by following  the prescription given in
\cite{mahan}, for the retarded Green function, which includes the
change of variables (recalling that $\vec z\equiv(z,\tau)$ )
\begin{equation}
i\tau\rightarrow -v_S t, \ \ \ \ \ \ \ \ \ \ \ \ i\omega\rightarrow
\frac{\omega}{v_S}+i\delta , \label{5.2.12}
\end{equation}
and the limit $\delta \rightarrow 0$. The soliton conductivity,
then, is given by
\begin{equation} \sigma_s=\lim_{\delta\rightarrow
0}\lim_{\omega\rightarrow 0}\lim_{k\rightarrow
0}\frac{1}{\omega}\textrm{Im}\left[\langle J
J\rangle_{ret}(\omega,k)\right]. \label{5.2.20}
\end{equation}
Taking the above limits, we obtain, after some algebra
\begin{eqnarray}
\sigma_s=\mathcal{Z}^{-1}\frac{N^2(\hbar
v_S)^2}{4\rho_0^4}\sum_{m=1}^{\infty}\frac{\left(\frac{\gamma\rho_0^N}{\hbar
v_S}\right)^{2m}}{(m!)^2}\int\prod_{i=1}^{2m}(iv_S\,dt_i)(dz_i)
\nonumber \\ \times \exp\left\{-\frac{N^2\hbar
v_S}{4\rho_0^2}\sum_{i=1}^{2m}\lambda_i\sum_{j=1}^{2m}\lambda_j
G(\vec z_i-\vec z_j)\right\}\nonumber \\ \times
\left\{\sum_{j=1}^{2m}\lambda_j G(\vec
z_j)\right\}\left\{\sum_{i=1}^{2m}\lambda_i\, t_i \right\}.
\label{5.2.23}
\end{eqnarray}

The non-transverse part of (\ref{5.2.10}) is a non-physical
``zero-point'' term, which must be subtracted from the current
correlator. Anyway it would not contribute to the conductivity
because it is real.

The temperature dependence of the soliton conductivity may now be
obtained by the usual methodology, through which we are led to the
version of (\ref{5.2.23}) having  finite integration regions
$0<\tau_i<\beta$ ($\beta=\hbar v_S/k_B T$) in the Euclidian time,
\begin{eqnarray}
\sigma_s^E(T)=i\sigma_s(T)\nonumber \\ =\frac{-i\mathcal{Z}^{-1}}{v_S}\frac{N^2(\hbar
v_S)^2}{4\rho_0^4}\sum_{m=1}^{\infty}\frac{\left(\frac{\gamma\rho_0^N}{\hbar
v_S}\right)^{2m}}{(m!)^2}\int_0^{\frac{\hbar v_S}{k_B
T}}\int_{-\infty}^{\infty}\prod_{i=1}^{2m} d\tau_i\,dz_i\nonumber \\
\times \exp\left\{-\frac{N^2\hbar
v_S}{4\rho_0^2}\sum_{i=1}^{2m}\lambda_i\sum_{j=1}^{2m}\lambda_j
G_T(\vec z_i-\vec z_j)\right\}\nonumber \\ \times \left\{\sum_{j=1}^{2m}\lambda_j
G_T(\vec
z_j)\right\}\left\{\sum_{i=1}^{2m}\lambda_i\,\tau_i\right\}.
\label{5.2.24}
\end{eqnarray}
In the above expression, the free massless Green function has been
replaced by the corresponding function at a finite temperature $T$,
namely, $G_T(\vec z)$ ($\vec z\equiv(z,\tau)$). This is a natural
consequence of the frequency quantization in the presence of a
finite interval for $\tau$.

The thermal Euclidian Green function of the 2D free massless scalar
theory in coordinate space has been evaluated in \cite{delepine} and
is given by
\begin{equation}
G_T(\vec
z)=-\frac{1}{4\pi}\ln\left\{\frac{\mu_0^2\,\beta^2}{\pi^2}\left[\cosh\left(\frac{2\pi
k_B T}{\hbar v_S}z\right)-\cos\left(\frac{2\pi k_B T}{\hbar
v_S}\tau\right)\right]\right\}. \label{5.2.25}
\end{equation}

Furthermore, the one-dimensional electrical conductivity is related
to the quantum soliton conductivity presented in (\ref{5.2.24}) by
\begin{equation}
\sigma_e=\left(\frac{e^2 v_S}{\hbar}\right)\sigma_s. \label{5.3.22}
\end{equation}
In order to obtain a result that could be compared with experimental
data, i.e., the three-dimensional electrical conductivity, we must
divide the above expression by the cross-section area of the
polyacetylene fibers, namely, $A\simeq\pi.10^4\, \textrm{{\AA}}^2$
\cite{shirakawa}. Then, we have
\begin{equation}
\sigma=\frac{\sigma_e}{A}. \label{5.3.23}
\end{equation}

In what follows, we will explicitly demonstrate that Eq.
(\ref{5.2.24}) yields an exactly vanishing quantum soliton
conductivity. First of all, let us change our notation by defining
\begin{equation}
\vec z_i= \begin{cases} \vec z_i^{\,+}\equiv (z_i^+,\tau_i^+), &
\textrm{for}\ \ 1 \leq i\leq m ; \\
& \\
\vec z_i^{\,-}\equiv (z_i^-,\tau_i^-), & \textrm{for}\ \ m+1 \leq
i\leq 2m .
\end{cases}
\label{novoz}
\end{equation}

Using the above notation and making the change of variables
$\tau_i^{+(-)}\rightarrow \left(\frac{2\pi k_B T}{\hbar
v_S}\right)\tau_i^{+(-)}$, we may then rewrite Eq. (\ref{5.2.24}) as
\begin{eqnarray}
\sigma_s^E(T)=i\sigma_s(T)=\frac{-i\mathcal{Z}^{-1}}{v_S}\frac{N^2(\hbar
v_S)^2}{4\rho_0^4}\sum_{m=1}^{\infty}\frac{\left(\frac{\gamma\rho_0^N}{\hbar
v_S}\right)^{2m}}{(m!)^2}\left(\frac{\hbar v_S}{2\pi k_B
T}\right)^{2m+1}\nonumber\\ \times\left(\prod_{i=1}^{m}\int_0^{2\pi}d\tau_i^+
\int_0^{2\pi}d\tau_i^-\right)\Xi_{\,m}(\tau_1^+,\ldots,\tau_m^+;\,\tau_1^-,\ldots,\tau_m^-),
\label{alex1}
\end{eqnarray}
where
\begin{eqnarray}
\Xi_{\,m}(\tau_1^+,\ldots,\tau_m^+;\,\tau_1^-,\ldots,\tau_m^-)\nonumber \\=
\left\{\sum_{i=1}^{m}\left(\tau_i^+-\tau_i^-\right)\right\}
\left(\prod_{i=1}^{m}\int_{-\infty}^{\infty}dz_i^+\int_{-\infty}^{\infty}dz_i^-\right)
\nonumber \\
\times \exp\left\{-\frac{N^2\hbar v_S}{4\rho_0^2}\sum_{i,j=1}^{m} \left[G_T(\vec z_i^{\,+}-\vec
z_j^{\,+})+G_T(\vec z_i^{\,-}-\vec z_j^{\,-})\nonumber\right .\right . \\ \left. \left .-G_T(\vec
z_i^{\,+}-\vec z_j^{\,-})-G_T(\vec z_i^{\,-}-\vec
z_j^{\,+})\right]\right\}\nonumber \\ \times\left\{\sum_{j=1}^{m}
\left[G_T(\vec z_j^{\,+})-G_T(\vec z_j^{\,-})\right]\right\},
\label{alex2}
\end{eqnarray}
in which, after re-scaling $\tau$,
\begin{equation}
G_T(\vec
z)=-\frac{1}{4\pi}\ln\left\{\frac{\mu_0^2\,\beta^2}{\pi^2}\left[\cosh\left(\frac{2\pi
k_B T}{\hbar v_S}z\right)-\cos \tau\right]\right\}.
\label{5.2.25new}
\end{equation}

We will now show that the quantum soliton conductivity vanishes
exactly. For this, we make the change of variables
$\tau_i^{+(-)}\rightarrow 2\pi-\tau_i^{+(-)} $ in the
$\tau$-integrals in (\ref{alex1}). Since
\begin{eqnarray}
\Xi_{\,m}(2\pi-\tau_1^+,\ldots,2\pi-\tau_m^+;\,2\pi-\tau_1^-,\ldots,2\pi-\tau_m^-)\nonumber\\
=-\Xi_{\,m}(\tau_1^+,\ldots,\tau_m^+;\,\tau_1^-,\ldots,\tau_m^-),
\label{alex5}
\end{eqnarray}
the announced result immediately follows.


\section{Conclusion}

The application of a general method of soliton quantization, based
on order-disorder duality, to a $Z(2)$ symmetric complex extension
of the effective field theory for the dimerization field of the TLM
model for polyacetylene has yielded an exactly vanishing result for
the quantum soliton conductivity. This strongly suggests that
dynamic solitons are not the charge carriers in polyacetylene in the
intermediate doping regime. The natural candidates are polarons.
However, as it happens in the case of solitons, which were studied
in the present work, a full quantum treatment is required in order
to derive a reliable expression for the polaron conductivity as a
function of the temperature. We are presently investigating the
quantum polaronic conductivity in this system.



This work has been supported in part by CNPq and FAPERJ. LM and AAS
were supported by CNPq. ECM was partially supported by CNPq.


\end{document}